\documentclass[review]{elsarticle}

\usepackage{lineno,hyperref}
\usepackage{lscape}
\usepackage{slashbox}

\linenumbers
\usepackage{siunitx} 

\newcommand{\angstrom}{\mbox{\normalfont\AA}}
\journal{Remote Sensing for Environment}

\biboptions{authoryear}

\begin{document}

\begin{frontmatter}

\title{Colour remote sensing of the impact of artificial light at night (II): Calibration of DSLR-based images from the International Space Station}

\author[1,2,3,4]{Alejandro S\'anchez de Miguel}\corref{mycorrespondingauthor}
\cortext[mycorrespondingauthor]{Corresponding author}
\ead{alejasan@ucm.es}
\author[2]{Jaime Zamorano}
\author[3]{Martin Aub\'e}
\author[5]{Jonathan Bennie}
\author[2]{Jesús Gallego}
\author[2,6]{Francisco Ocaña}
\author[7]{Donald R. Pettit }
\author[8]{William L. Stefanov}
\author[1] {Kevin J. Gaston}
\address[1]{Environment and Sustainability Institute, University of Exeter, Penryn, Cornwall TR10 9FE, U.K. }
\address[2]{Depto. F\'isica de la Tierra y Astrof\'isica. Instituto de Física de Part\'iculas y del Cosmos (IPARCOS), Universidad Complutense, Madrid, Spain}
\address[3]{Physics dept., CEGEP de Sherbrooke, Sherbrooke, J1E 4K1, Canada}
\address[4]{Instituto de Astrof\'isica de Andaluc\'ia, Glorieta de la Astronom\'ia, s/n,C.P.18008 Granada, Spain}
\address[5]{Centre for Geography and Environmental Science, University of Exeter, Penryn, Cornwall TR10 9FE, UK}
\address[6]{Quasar SR for ESA, European Space Astronomy Centre, E-28691 Villanueva de la Cañada, Spain}
\address[7]{Astronaut Office, Flight Operations Directorate, NASA Lyndon B. Johnson Space Center, Houston, Texas, USA}
\address[8]{Exploration Science Office, Astromaterials Research and Exploration Science Division, Exploration Integration and Science Directorate, NASA Lyndon B. Johnson Space Center, Houston, Texas, USA }


\begin{abstract}
Nighttime images taken with DSLR cameras from the International Space Station (ISS) can provide valuable information on the spatial and temporal variation of artificial nighttime lighting on Earth. In particular, this is the only source of historical and current visible multispectral data across the world (DMSP/OLS and SNPP/VIIRS-DNB data are panchromatic and multispectral in the infrared but not at visible wavelengths). The ISS images require substantial processing and proper calibration to exploit intensities and ratios from the RGB channels. Here we describe the different calibration steps, addressing in turn Decodification, Linearity correction (ISO dependent), Flat field/Vignetting, Spectral characterization of the channels, Astrometric calibration/georeferencing, Photometric calibration (stars)/Radiometric correction (settings correction - by exposure time, ISO, lens transmittance, etc) and Transmittance correction (window transmittance, atmospheric correction). We provide an example of the application of this processing method to an image of Spain.
\end{abstract}

\end{frontmatter}


\section{Introduction}\label{S:1}

There is growing demand for colour imagery of the Earth at night. This has particularly been driven by increasing recognition of the impacts of outdoor artificial nighttime lighting (from streetlights and other sources) on the natural environment, human health, and associated policy and public concerns (e.g. \citet{rich2013ecological}; \citet{holker2010dark}; \citet{falchi2011limiting}; \citet{gaston2012reducing,gaston2015benefits}; \citet{gaston2013green,gaston2018lighting}; \citet{garcia2018evaluating}). These impacts are not only influenced by the spatial extent, the timing and the intensity of that lighting, but also by its spectrum. This last consideration limits the insights that can be obtained from the vast majority of historical and current sources of geographic scale data on the occurrence of such lighting, including data from the Defense Meteorological Program/Operational Line-Scan System (DMSP/OLS; \cite{elvidge1997mapping}) and from the Suomi-National Polar-Orbiting Partnership/Visible and Infrared Imaging Radiometer Suite - Day/Night Band (SNPP/VIIRS-DNB; \cite{elvidge2013viirs}).

By far the most important source of colour imagery of the Earth at night is that obtained from the International Space Station (ISS) by the astronauts and cosmonauts of the five space agencies that constitute the ISS consortium: NASA, ESA, JAXA, CSA-ASC and ROSCOSMOS (\citet{stefanov2017astronaut,stefanov2019astronaut}). Between 2002 and 2018, more than 3 million images have been taken by the crew of the ISS, of which 1.34 million were taken at night (sun elevation less than 0 degrees). The diurnal and nocturnal images from NASA, CSA-ASC and ESA are stored in the NASA archive (http:/eol.jsc.nasa.gov; \citet{stefanov2017astronaut,stefanov2019astronaut}). Some, but not all, of the images taken by JAXA and ROSCOSMOS are also stored in the NASA archive. The scientific value of these images is immense, as they constitute the only large public dataset with nighttime colour information obtained from space over the last 17 years (Levin et al., 2020). Due to the limitations of data from the VIIRS and DMSP satellites, this dataset is key to determining change in visible artificial lighting over this long period (\citet{kyba2017artificially}; \citet{sanchezde2019colour}).

The camera bodies and lenses used by the astronauts to take images of the Earth have mainly been from Nikon’s professional product line (see Table 1 for those used for nocturnal photography). These have usually been unmodified, with the exception of the Nikon D3S S/N:2007944 used on missions ISS030 and ISS031 that had the infra-red filter removed. Most of the Nikon cameras used have had a CMOS sensor, although the Kodak 760C and the Nikon D1 that were used had CCD sensors (\citet{waltham2013ccd}). All the cameras have a Bayer filter that provides simultaneous images in three colours (two green, one red and one blue).

The environmental conditions on the ISS are strictly controlled, with temperature at 24ºC $\pm$ 2ºC, pressure around 745-721 mmHg, and relative humidity around 60\%. This means that the cameras have not only been operating well within their nominal environmental ranges but effectively in constant environmental conditions. Although cameras have been used during extravehicular activity (EVA) or in external experiments, principally windows in the cupola and elsewhere on the ISS enable astronauts to take images of the Earth from a wide variety of angles. The ground track of the ISS covers nadir latitudes from 51.6º N to 51.6º S, at an altitude of approximately 405 km, although this can vary in the range of 330 to 435 km.  Usually images have been taken by astronauts using cameras that are handheld. \citet{pettit2009exploring} developed a prototype system that partially compensated for movement during image acquisition and was used during mission ISS006. Later missions, like ISS030 to ISS040, have employed a special tripod (Nightpod; \citet{sabbatini2014nightpod}). At least for images held in the open NASA archive there is no information on their acquisition beyond the metadata of the images themselves, thus it is not easy and frequently impossible to attribute them to a specific window, tracking device or operator. Only when an astronaut posts their own images, on Twitter for example, can we have reasonable confidence in their authorship.

A growing number of science studies have attempted to use images taken from the ISS in their raw or crudely enhanced forms \citep{venzke2009reports,lockwood2013volcanoes,anderson2010characterizing,levin2012high,liu2011relationships,metcalf2012detecting,kuechly2012aerial,mazor2013can,sanchez2014atlas,li2014quantifying,so2014observational,kyba2014high,rybnikova2017remote,rybnikova2018population,xu2018mapping,levin2020remote}. Indeed, the only research work we are aware of that made some attempt to address issues of calibration is that of \citet{so2014observational}, which used a dark subtraction, and those in which the current team has been involved \citep{zamorano2011iss,sanchezde2015variacion,garcia2018evaluating,garcia2019artificial,hale2019artificial,salvador_bara_2020_3981095,pauwels2019accounting,sanchezde2019colour,sanchezde2020nature}. The effects of this lack of calibration can be very different from study to study and need to be considered carefully. Unfortunately, in many cases calibration of images taken by astronauts from the ISS will be necessary in order to provide accurate representations of the colour composition of scenes, and this is not a trivial task.

In a previous paper \citep{sanchezde2019colour}, we described the use of colour-colour diagrams to analyse images taken by astronauts on the ISS and to estimate spatial and temporal variation in the spectrum of artificial lighting emissions. In this paper we provide a methodology for calibrating such images. The approach that we describe is also relevant to the calibration of images taken with standard DSLR cameras for other purposes, including meteor science (meteor photometry - \cite{borovivcka2014spectral}; meteor video photometry - \cite{madiedo2019multiwavelength}; meteor spectroscopy - \cite{chengcheng2011}), measurement of skyglow \citep{hanel2017measuring}, and more generally scientific photography using DSLR cameras with antiblooming. In order to use well characterized standard emitters, we use stars as calibration sources, as the variability and stability of their spectra is well known. Indeed, laboratory calibration sources typically have a precision of 0.2\% \citep{Wolfe:1613508}, whilst calibration by stars can attain precisions of 0.001\% \citep{poddany2010exoplanet}. This technique also allows us to use multiple calibrations from onboard the ISS that are stable in time and that account for the window through which images were obtained. A similar technique has also been proposed for calibration of data from the SNPP/VIIRS-DNB \citep{fulbright2015suomi} and was suggested first by \cite{zamorano2011iss}.

Calibration of images of the earth acquired from the ISS requires a number of steps that we will address in turn (Fig. 1): Decodification, Linearity correction (ISO dependent), Flat field/Vignetting, Spectral characterization of the channels*, Astrometric calibration*/georeferencing, Photometric calibration (stars)*/ Radiometric correction (settings correction - by exposure time, ISO, lens transmittance, etc) and Transmittance correction (window transmittance, atmospheric correction). The steps marked with an asterisk may not be required for some specific science cases (e.g. differential evolution time - \cite{meier2018temporal,bara2019estimating};  edge detection - \cite{kotarba2016impervious}). Neglecting any particular correction should always explicitly be justified. 

In this paper, we will focus primarily on calibration of images taken with the Nikon D3S (for calibration equipment see Supplementary Information) because this is the best studied case and the camera that has been used most intensively for nocturnal photography at the ISS over the two last decades (from mission ISS026 until mission ISS045). However, in general, with small adjustments, the procedures described can be applied to any other digital camera, and we will consider all of the cameras used at the ISS (Table 1) and highlight the differences between them when it is relevant. The Nikon D3S has a 12.1 megapixel sensor equivalent to a full frame (35mm) and was announced by Nikon Corporation on 14 October 2009. It has interchangeable lenses using the F-mount, an ISO range from 200 to 12800, and 14-bit A/D conversion.

\section{Calibration steps}
\subsection{Step 1: Decodification}

A DSLR camera obtains information as photons of light that produce electrons, and these electrons are effectively stored in pixels of the CCD/CMOS chip. The electrons are trapped by the nearest potential well (one per pixel) and digitalized by an analogue conversion (DA Converter) sensor that measures the electrical potential on the wells of the chip. Whilst in some professional cameras a non-destructive read out is possible \citep{nakamura1995cmos}, most CCD/CMOS chips are read by measuring the current when it passes though the DA converter \citep{fowler1995cmos,fossum2000active}. This information is pre-processed and saved as a data file. This file is usually coded in a proprietary format, and in the case of Nikon cameras, this is called NEF. This is what is called a RAW file. The first step in the calibration process is the decodification of this file, taking the data coded in the proprietary format and transforming it into a readable open format, such as TIFF \citep{desk1986tiff} or FITS \citep{hanisch2001definition}, without losing crucial information or changing the physical meaning.

Even so-called RAW files from a DSLR tend to have undergone some pre-processing. A pure ideal CCD/CMOS RAW file coming from an ideal sensor will have three components: (i) the signal itself, (ii) an artificial signal (bias) included on the sensor to avoid negative values in the process of digitalization, and (iii) Gaussian noise plus an additive constant (dark current) arising from the sensor’s sensitivity to heat. On our reference camera, the Nikon D3s, and all subsequent versions except the Nikon D5, a complex hardware treatment is applied to the signal \citep{koyama2011development,sanchezde2015variacion}. This treatment means that we do not need to do any bias or dark correction in these cases, but noise effects need to be controlled. DSLR sensors are capable of a 16-bit dynamic range. However, in practice on commercial models it is 14 or 12 bits, because saving 16-bit information takes too much time. This means that some detailed information is lost in the digitalization process with a loss of dynamic range.

The current most widely used software to undertake decodification is DCRAW \citep{coffin2008dcraw} and the most conservative decodification options are: -6 that means 16 bit extraction, -o 0 that means no color balance, -H 1 that means nothing excluded, and -v to show any error in the decodification process.

\subsection{Step 2: Linearity correction}
While most scientific camera sensors force linearity between numbers of incident photons and pixel signal strength as a key characteristic, commercial DSLR cameras have sensors with an anti-blooming function. By using an overflow channel, this system ensures that when a well of the chip is saturated then excess electrons do not contaminate adjacent wells (as can happen with regular CCD or CMOS sensors)\citep{sakai2009pixel}. This produces a roughly linear response in the pixel signal when high gain is used. Unfortunately, there are no details available of the specific technology used for the Nikon DX sensors. Currently, we have not detected differential absorption by chip substrate, although this could be investigated further in future versions of the processing pipeline.

Typically, the gain of the camera operates to an ISO standard \citep{ISO12232}. Fig. 2 shows the effect of anti-blooming when a camera is exposed to a stable light source across all of the different exposure times that the camera is able to produce. This effect can be compensated by characterization of this response for a particular ISO. Even if a camera is in principle calibrated to the ISO standard, this still allows a 20\% error, and there can be differences on the different channels. Indeed, without corrections, bad colour information (aka. band ratios, see \cite{sanchezde2019colour}) can be produced. As \cite{metcalf2012detecting} found, radiance values can be underestimated, and radiance ratios can appear variable even if they are stable.

Having characterised the response of the camera for a particular ISO, the intensity values of an image that has been taken using this can be divided by these corrected ones. Those over the range of the linear response will not change (aka. under $\sim17.000$ ADU)  whilst those in the range of the non-linear response will be linearized.

\subsection{2.3 Step 3: Flat field}
A wide range of lenses have been used with DSLRs on the ISS, from fish eyes to telephotos. A few have been used for long periods, but others on just a single mission. These lenses introduce various additional calibration issues. The first of these is flat field/illumination/vignetting correction. This is a very standard correction with DSLR imagery, although unfortunately not usually addressed with imagery from the ISS (see \citep{burggraaff2019standardized} for standard corrections to consumer cameras). This correction consists of the acquisition of shots from a uniform emitting source (aka. flat field), so that any heterogeneity in the acquired image is the result of vignetting of the lens and linearity effects. Distortion effects, such Barrel, Pincushion, and Mustache can also be corrected. However, this should be avoided unless absolutely necessary because of the challenges of combining such corrections with those necessary due to the angular perspective at which many ISS images are obtained as a consequence of the Earth’s curvature, and because performing several successive distortions will amplify errors \citep{cardiel2002proper}) (see 2.7 Step 6).

Images of star fields from the ISS are not available for all combinations of cameras and lenses that have been used to obtain images of the Earth. In order to translate the light intensity measured using one lens to that which would be measured using another the light transmission of the optical elements of the latter needs to be considered. The f/ number can give a first order idea of this in a standardised way. However, this is not sufficiently accurate for many purposes, so instead the Transmission coefficient (T number) of the lens needs to be measured. Table 2 provides these values, as well as the f/ number for some of the most frequently used lenses on the ISS. Values in this table have been calculated using the SaveStar Consulting S.L. lab by acquiring images of a $4 \pm 0.1$ lux illuminated lambertian surface. The light Source was a tungsten filament.

Note that in general the apertures of lenses can be closed down to provide sharper images.  When this happens, the diaphragm rarely closes to exactly the same position (except when fully open), so it is recommended that imagery is taken using maximum apertures (explained in more detail in 2.5 Step 5c). On bright sources this can be considered a compromise between sharpness of images, signal to noise ratios, and photometric accuracy. Fortunately, this has been done for most of the images taken at night from the ISS.

\subsection{Step 4: Astrometric calibration}

The astronauts do not only acquire images of cities at night. They also acquire lots of other kinds of nocturnal images, including   of auroras, sunsets, and occasionally also star fields.  The "Cities at Night" project has a NASA archive selection of these images\citep{citiesatnight}. They can be used for several proposes, such as calculating lens distortion and lens transmittance, or as we do in this case, for radiometric calibration.

Once the characteristics of a camera and lens have been determined it is necessary to use standard sources to calibrate the imagery of the Earth that has been taken. The use for this purpose of starfield images taken from the ISS has the great advantage that they were obtained with the equipment under the same temperature, pressure and humidity conditions as the images of the Earth. They will also have been acquired through windows with the same reflection and transmission characteristics; light transmission through the windows is very high except for some Zvezda windows (we can assume that absorption in all bands is less than 5\% in the visible regime, although windows in the Destiny lab show significant transmission reduction beyond the NIR), and special coatings have been used to avoid reflections (see 2.9 Step 8). 

The first step in using starfield images as standard sources is astrometric calibration, determining the coordinates of the stars in each image. This can be done using standard astronomical methods. We use the software Astrometry.net \citep{lang2010astrometry}. This automatically spatially calibrates an image, so that each pixel has corresponding celestial coordinates. Moreover, it extracts the sources and identifies them using a catalog. Stars are the flux standard sources that we will subsequently use.

Astrometric calibration (also called the World Coordinate System) by Astrometry.net provides direction, orientation and plate scale (transformation between the apparent angular separation and linear separation at the focal plane). The software does not require any additional input, and makes a blind calibration using only the image. It is a powerful tool as there is no need to know spacecraft attitude nor window orientation nor the direction in which an astronaut was pointing the camera. It has also been used successfully in non-stabilised high-altitude balloon observations where attitude is not controlled \citep{ocana2017techniques,ocana2019balloon}.

\subsection{Step 5a: Photometric calibration on board}

To perform photometric calibration, we need a standard source. As there are no electronic standard flux sources onboard of the ISS the most accurate method of calibration is using the stars, as has been proposed for other satellites \citep{fulbright2015suomi} and as is routinely done in astronomy.
  
We will focus on use of a selection of stars from \cite{ducati2002vizier}. Even though stars are very well characterized light sources, there are differences between the filters that were used in making those measurements and those used in DSLR cameras. To solve this problem we can use a statistical correction based on the synthetic photometry technique \citep{straizys1996method,sanchez2017sky,sanchezde2019colour}:
\begin{equation}\label{1}
GN = V + 0.1291(2) \times(B-V) -0.0051(2)
\end{equation}
\begin{equation}\label{2}
BN = GN + 0.6123(7) \times (B-V) -0.0340(6) 
\end{equation}
\begin{equation}\label{3}
RN = GN + 0.0262(3) + 0.5880(5) \times (R-V)
\end{equation}
where BN, RN and GN are the respective RGB bands of the DSLR image, and B, V and R are Johnson bands (an astronomical standard); these relationships are shown in Fig. 3. These corrections have been calculated for the spectral response of the D3S although, as can been seen in \cite{sanchezde2019colour}, all Nikon cameras have extremely similar spectral responses. Once these corrections to the intensities of the stars are applied, the transformation from data numbers to radiometry units is immediate by applying a linear fit (see Fig. 4).

As the \cite{ducati2002vizier} catalogue has few stars in the dim part of the calibration and only 18 stars in total, for verification purposes we used the Tycho catalogue \citep{hog2000tycho}, which contains many more stars.Fig. 5 shows the effect of the saturation of the brightest stars. Future analysis can be done the new catalog \cite{cardiel2021synthetic}, specifically designed for DSLR cameras. 

By using these methods we get the correspondence between the radiance and the data numbers in magnitudes. To convert this on the international system of units we need to use:
\begin{equation}\label{4}
AB = -2.5 \log_{10}(flux) -5 \log_{10}(w) +2.401
\end{equation}
where AB is in $mag / arcsec^{2}$,
flux in $nW / cm^{2} / sr / \si{\angstrom}$,and w is wavelength in $\si{\angstrom}$ \citep{sanchez2017sky}.

When we calibrate a starfield image taken from the ISS, we do not need to apply any atmospheric correction, as is usually done when this technique is used with starfield images taken from the ground, because the ISS is at 400 km above the Earth and the density of the atmosphere there is negligible and for this purpose is considered to be a vacuum (at this elevation pressure is even higher than that of the regular vacuum labs on Earth, from $10^{-6}$ to $10^{-9}$ torr \citep{finckenor2017researcher}, compared to the 760 torr of the atmosphere at sea level).

\subsubsection{Step 5b: Photometric calibration from the ground}
As previously mentioned, some images of the Earth taken from the ISS have been obtained using lenses for which starfield images are not available (for example, because their focal lengths make it impossible to take sharp starfield images given the speed of movement of the ISS). For those cases when starfield calibration is not possible, the solution is to apply a standard absolute photometry technique to achieve photometric calibration of the lens from the ground.

For observations taken from the ground the light of the stars travels through the atmosphere and suffers absorption, scattering, and dispersion depending on the wavelength. The amount of extinction (the combined effect) depends on the extinction coefficient of the atmosphere at this time and on the length of the path of the light across the atmosphere. As a result, the flux density (in units of $ergs / s / cm^{2} / \si{\angstrom}$) measured on the ground for a star is lower than the flux at the top of the atmosphere:
\begin{equation}\label{5}
F \mbox{(observed)} = F_{o}\cdot 10^{-0.4 K X}
\end{equation}
where $F_{o}$ is the flux outside the atmosphere (which we know since we are observing standard stars), K is the extinction coefficient for this wavelength, and X the airmass calculated as X = sec z (this formula is only usable for zenith angles up to about $60^{\circ}$ to $75^{\circ}$; for more accurate version see 2.8 Step 7), z being the zenith angle (90 - elevation above horizon). We call photometric nights those clear nights with constant transparency (see definition of photometric night at https://www.eso.org/sci/observing/phase2/ObsConditions.html). For these nights the extinction coefficient is constant for the whole night. Observing several standard stars through the night at different heights above the horizon we can derive this coefficient and the zero point of the photometry for each photometric band. The derived zero point of the photometric band is valid for this observational setup (camera, lens, ISO, and f number) regardless of the night of the observation, i.e. it depends only on the instrumentation. So, the atmospheric effects only need to be considered on the particular night that the calibration images are acquired.

\subsubsection{Step 5c: Setting adjustments}
It is not possible to have images calibrated for all of the possible cameras and settings. So, we scale the adjustments to 50 mm f/1.4. It is well known that the repeatability of mechanical iris-type lens diaphragms is limited, having a noticeable tolerance. The blades move back and forth during each image acquisition, limiting the entrance pupil to approximately the same position but not exactly so \citep{chylinski2012time}, creating so-called ‘aperture flickering’. We have created a model to estimate the potential effect of closing the shutter but the current one is a general correction with the f/ number:
\begin{equation}\label{6}
L_{0} = 2 \times f/^{-2} \:\mbox{ or }\: Tn 
\end{equation}

\begin{equation}\label{7}
\mbox{Correction factor} = 1/(ISO/100) \times C_{0}/T/B_{N} \times C_{1}/L_{0} 		
\end{equation}
where, ISO is the ISO 12232:2006 standard for digital photography, $C_0$ is a correction for the sensitivity of the camera model, T is the exposure time, $B_N$ is the correction for the bit rate, and $C_1$ is the colour correction between different camera models. $L_0$ is the correction for the aperture expressed as the f number or the Tn true transmission of the lens when the shutter is fully open (see Table 2). Some of these settings can have up to 15\% error according to the ISO 12232:2006 standard.

\section{Step 6: Georeferencing}
Images of the earth from the ISS need to be georeferenced to establish the link between image pixels and actual points on the ground. The time of acquisition of each image is known but, although the orbit of the ISS is precisely defined, unfortunately this does not provide sufficient information. The time stamp data are not precise enough to define a nadir location (errors may be more than 1000 km), and an image may be taken in a direction that departs substantially from the nadir. A major citizen science program, “Cities at Night” (http://www.citiesatnight.org,\citep{de2014atlas}), is being used to automate the georeferencing of large numbers of ISS images by identifying the urban areas that these represent. More generally, and where sufficient control points cannot automatically be identified, georeferencing can be done manually, for which purpose we have used QGIS \citep{qgis2015qgis} and Global Mapper \citep{geographics2011global}. Several Python libraries have been used in this process, like NumPy \citep{oliphant2006guide}, AstroPy \citep{robitaille2013astropy} and GDAL \citep{warmerdam2008geospatial}. The preferred sampling method is bilinear for the pixels and thin plate spline for the coordinates. The number of recommended control points can depend a lot on the inclination of the image and lens. For nadir and 400 mm, 20 control points can be enough to reach RMSE 2.5 pixels using polynomial third degree fitting. For tilted images taken with a 24 mm, for example, more than 100 points would be needed and with 400 points can be visually indistinguishable from a comparison layer (VIIRS); the only transformations able to rectify deformations on tilted images are Thin Plane Splines that provide unrealistic values of RMSE - a conservative estimation can be RMSE $\sim 4$ pixels, but more research is needed to systematise this analysis.

\subsection{Step 7: Atmospheric correction}
Whilst no atmospheric correction is required for starfield images taken from the ISS, this needs to be done for images taken of the Earth. The procedure selected depends on the concept of airmass. In astronomy, one airmass is equivalent to the volume of atmosphere in the direction of the zenith of an observer on the ground (that is equivalent to the same volume of mass to an observer located in orbit looking to the nadir at the same geographical coordinates). Using this approach it is straightforward to estimate the equivalent volume of atmosphere at different zenith angles or the corresponding nadir angles. The formula X = sec z, as above, can be used up to 80 degrees of zenith angle considering a plane parallel approximation of atmosphere with about 3\% error and for higher angles several models are available (e.g. \citet{pickering2002southern}). Then, we can atmospherically correct each image using:

\begin{equation}\label{8}
I/I_{0} = 10^{X_h(h)\times K/-2.5} 		
\end{equation}

where $X_{h}$ is the airmass function of the height, h is the height (in metres), I is the observed intensity, $I_{0}$ the intensity with atmospheric effect, and K is the extinction coefficient (derived from \cite{Harwit1973}). The extinction coefficient for 1 air mass can be calculated as:
\begin{equation}\label{9}
K=A_{rag}+A_{aer}+A_{oz}
\end{equation}
where $A_{rag}$ is Raleigh scattering, $A_{aer}$ the absorption by aerosols, and $A_{oz}$ the absorption by ozone. Using the definitions compiled in \cite{garcia2012contaminacion}:

\begin{equation}\label{10}
A_{ray}=9.4977\times10^{-3} \left(\frac{1}{\lambda}\right)^{4} c^{2} \times \mbox{exp} \left(-\frac{H_{obs}}{7.996}\right)
\end{equation}

\begin{equation}\label{11}
c = 0.23465 +\left( \frac{1.076 \times 10^{2} }{146 - (1/\lambda^{2})}\right) + \left(\frac{0.93161}{41 - (1/\lambda^{2})}\right)
\end{equation}

where $H_{obs}$ is the height of the observer, $\lambda$ is the wavelength in microns, and "c" is the air's refraction index.

\begin{equation}\label{12}
A_{aer} (\lambda,h) = A_{o}\lambda^{-\Sigma} exp(- H_{obs}/H)
\end{equation}
where H is the density scale height for aerosols and $A_{o}$ is the total optical thickness of atmospheric aerosols for $\lambda = 1 \mu m$, which depends on the total content of particles and on their efficiency for scattering and absorption and is taken to be 0.087 \citep{walker1988effect,mohan1999atmospheric}. $\Sigma$ is a parameter that depends on the size of the aerosol particles.

Each of these factors $(A_{rag}, A_{aer}, A_{oz})$ is wavelength dependent and also depends on atmospheric conditions. By convention, by default we have considered the aerosol content conditions of the AERONET \citep{holben1998aeronet} Madrid station on the 29th March 2012 (AOD340 = 0.333, AOD380 = 0.305, AOD440 = 0.251, AOD500 = 0.208, AOD675 = 0.127, AOD870 = 0.084, AOD1020 = 0.066), interpolated for the center of the Nikon bands. We use \cite{rayleigh1899xxxiv} to consider the Rayleigh scattering (equation 10), \cite{stalin2008night} to consider the aerosols (equation 12) and \cite{hayes1975rediscussion} to consider the Ozone absorption ($A_{oz}$ factor). To correct for all of these effects equation 8 should be applied to each pixel of the image. Currently, the aerosol content correction is only accurate for a narrow field of view, for a wide field of view different AODs would be needed per pixel. In the future, these data could be obtained from VIIRs products \citep{huang2016validation}. Currently, no other nocturnal lights products provide multispectral aerosol corrections and the only other nocturnal light product that provides some correction of this nature is the dimensional panchromatic aerosol corrections from VIIRS VNP46A2 \citep{roman2018nasa} that are still in the beta phase.

To determine the original location of the ISS we use a photogrammetric technique. Considering a circle on geographical coordinates, we will obtain an ellipse on the raw image coordinates. The axis of the ellipse is either coincident with the direction of observation or perpendicular to the direction of observation. Matching those two directions with the ISS ground track, we obtain two potential solutions for the true nadir of the ISS. Most of the time the minor axis indicates the direction of the true nadir, although, because of lens aberrations and the curvature of the earth, when the ISS astronauts take images very close to the nadir it can be the major axis that will align with the true view direction (see Fig. 6). Most of the time, a visual inspection of the image allows the identification of the true nadir, by looking at the patterns of the streets and how the buildings block the light of the streets.

Once we know the true nadir, as the altitude of the ISS is also known, it is straightforward to determine the tilt angle (we define the tilt angle as that between the nadir and the view angle of the center of the image). Then, the atmospheric extinction can be determined using the approach used in astronomy because of the Helmholtz reciprocity law, the same extinction that an observer has looking to the ISS from a location A is what an observer on the ISS has looking to location A. As the lens used may have a very wide field of view, the distance from the ISS to the ground can be significantly different from one part of an image to another, so we calculate this extinction for each pixel of the image.

In some ISS images of the Earth at night, high thin clouds, fog, etc are apparent. These are challenging to address in the generic pipeline described here, and may require more of a case-by-case approach, employing available tools.  So, currently, we recommend discarding any areas where the Point Spread Function (PSF) of the emission sources is not homogeneous.

\subsection{Step 8: Intercalibration}
For a variety of reasons, there may be some systematic differences in the extrapolation of the calibration with stars. One of the main unknowns in analysing an image taken from the ISS is which window was used. Sometimes this can be determined because of the date on which the image was obtained or because the image was taken using the Nightpod, which is always used at the main Cupola window. The most popular windows are the Cupola, Window Observational Research Facility (WORF), and Window 7 of the Zvezda module. The Cupola and WORF window have very similar transmission characteristics and so, in principle, should not create differences in colour, although some ESA contacts have reported lower transmission of the WORF. This effect is explained mainly because of the Cupola scratch pane, that is not always in place as it is removable. Window 7 of the Zvezda module is comparable although it is even more transparent in principle as it is also UV and IR transparent \cite{Pettit2006night}; more information about the transmission can be found in \cite{ESA2011} and \cite{stefanov2017astronaut,Stefanov2019}. Transmittance can also vary greatly with the angle to the window at which an image was taken. It is typically impossible to know what this angle was, although the geometry of the Cupola, WORF, and Zvezda do not allow acquisitions with very shallow angles.   

Apart from the different windows used, another reason why there may be systematic differences between the extrapolation of the calibration with stars with different lenses is the use of shutters that are not fully open. 

For these reasons, and possibly others, we might want to intercalibrate different images or calibrate them against another source, such as data from SNPP/VIIRS-DNB. In a future paper we will focus on these techniques.

\section{Spain/Madrid: a worked application}
Artificial nighttime lighting has been more extensively studied across Spain than almost any other country \citep{sanchezde2014evolution,kyba2014high,estrada2016statistical,oriol2017,ges2018light,tong2020angular}. To provide a worked example of the calibration of an astronaut photograph, we use a wide angle image of Madrid and its environs (Fig. 7) taken from the ISS (ISS053-E-249189) on 19/11/2017 at 21:51:38 GMT (note: that the database time is incorrect by 1 h in this exceptional case. This is known because the nadir location that corresponds with the RAW EXIF information is in the Pacific Ocean [lat -49.2, -126.2], which is clearly not possible. The 1 h corrected location is just $\sim200$ km [lat 38.4,lon -6.4]) from the true nadir [lat 37.49,lon -7.77]) and downloaded on 21/05/2020 from NASA's Gateway to Astronaut Photography of Earth (https://eol.jsc.nasa.gov/). Fig. 8 shows the RAW image. This is extracted and separated into four different colour images (Fig. 9). These are not linear at the high values, so linearity correction is applied (Fig. 10). Fig. 11 shows the flat field correction that needs to be applied given that the original image was taken with a 24mm lens. The parameters of equation 7 are, according to the image metadata and camera model reference Nikon D3s, T=1/10, C0=1, C1=1 for the Green channel, 1.12 for the Red Channel and 0.95 for the Blue Channel, ISO=5000, BN=4, and Tn=1.050.

Photometric calibration was carried out for the colour images using the stars (Fig. 12), followed by georeferencing (Fig. 13). Once we have the transformation between geographic location and pixel, we can generate a map of the view angle of the camera with respect to the ground (Fig. 14), which gives the amount of atmosphere that the light goes through before hitting the sensor. After this correction we have a calibrated image that is internally radiometrically coherent and colour coherent (Fig. 15). This could now be compared with other ISS images and intercalibrated with them using reference points or compared with VIIRS images. 102 control points were used with an RSME of $\sim 4$ pixels (due to the large deformation of the image, the pixels closer to the nadir have higher precision, and points close to the horizon less. Also, the ISS sampling is 3 times higher than that of the reference from VIIRS, so it is natural that the image cannot have higher accuracy than the pixel size of the reference layer).

Fig \ref{fig:16} and Fig \ref{fig:17} show how this calibrated ISS image compares to a VIIRS image. The comparison is not simple as the sensors used have different spectral sensitivity, different flyby times and were obtained at different angles, and there are not sufficient calibrated ISS images yet fully to determine any systematic differences between the emissions detected in ISS and VIIRS data. fig.\ref{fig:18} shows that in comparing the calibrated ISS image of Madrid and VIIRS data the main differences are due to small georeferencing/resolution issues, and light sources that are visible on one image but not the other because of the time of acquisition or the tilt. Nonetheless, despite in the present example there being a 1  month difference in the timing of acquisition of the data, 4 h difference in the flyby time, spectral sensitivity difference, and difference in spatial resolution, there is predominantly a linear relationship between the emissions detected by the two images, and they compare very well. There is an absolute offset between the two of a factor 1.92. This could be explained by several factors, including dimming of the streetlights of Madrid after midnight.

\section{Madrid: ISS-ISS intercalibration}
In order to examine the performance of the calibration of ISS imagery, we selected four images of Madrid (see fig. \ref{fig:19}). All were acquired before a major street lighting change took place in the city in 2014 \citep{robles2021evolution}.

\begin{table}[ht]
\begin{tabular}{|l|l|l|l|l|l|l|l|}
\hline
File          & Tilt & Exp T. & f/  & ISO   & Lens          & Cam    & Time                   \\ \hline
iss030e292893 & 7.0  & 1/50   & 2.8 & 10000 & 180mm f/2.8 D & D3S & 2012:02:08 21:09:16.87 \\
iss030e292895 & 4.0  & 1/50   & 2.8 & 10000 & 180mm f/2.8 D & D3S & 2012:02:08 21:09:02.37 \\
iss031e095634 & 7.0  & 1/50   & 1.4 & 3200  & 85mm f/1.4 D  & D3S & 2012:06:04 22:26:49    \\
iss035e023371 & 29.0 & 1/40   & 3.2 & 51200 & 400mm f/2.8   & D3S & 2013:04:18 22:10:09.73 \\ \hline
\end{tabular}
\caption{\label{tab:images1}Specifications of the images used for comparison.}
\end{table}

The two first images were obtained a few seconds from each other, but with small changes of view angle and exactly the same camera settings, so they are probably as similar as images taken from the ISS are likely to be. The third image, was obtained with a different lens, ISO and aperture settings, and was acquired 1 h and 30 min later, but with a similar view angle. The fourth image was again obtained with a different lens, ISO, aperture and view angle, still in the same daily time frame, but more than one year apart. A 5 pixel Gaussian blur was applied to each image to minimize effects of errors in georeferencing.

R2 values for the pixel by pixel comparison of light intensities determined from the four images varied between 0.88 and 0.98 (Table 2), with the majority of data points lying close to the 1:1 line (Figure 20). The differences in intensities between images increase with variation in how those images were acquired. Given the challenges of nocturnal remote sensing that result from the nature of the light sources \citep{tong2020angular}, we conclude that these are satisfactory results and comparable to findings for panchromatic data from satellite sensors \citep{coesfeld2018variation}.

\begin{table}[ht]
\begin{tabular}{|l|cccc|}
\hline
\backslashbox{Ref}{Pro} & \multicolumn{1}{c|}{iss030e292893} & \multicolumn{1}{c|}{iss030e292895} & \multicolumn{1}{c|}{iss031e095634} & iss035e023371 \\ \hline
iss030e292893 & 1    & 0.98 & 0.95 & 0.88 \\
iss030e292895 & 0.98 & 1    & 0.96 & 0.88 \\
iss031e095634 & 0.95 & 0.96 & 1    & 0.88 \\
iss035e023371 & 0.88 & 0.88 & 0.88 & 1    \\ \hline
\end{tabular}
\caption{\label{tab:images2} $R^2$ values corresponding to the plots fig \ref{fig:20}. }
\end{table}

\section{Conclusion}
Calibration of nighttime images of the Earth taken by astronauts aboard the ISS is not, in general, an easy task. These images were never designed to provide remote sensing data. However, if handled properly, they can provide multispectral information on the distribution and change in nighttime lighting that is not available from any other source, and at spatial resolutions that are comparable or better than those obtained by the DMSP-OLS and SNPP/VIIRS-DNB platforms. Reflection correction is not needed and we did not apply it because most studies of light pollution are concerned with the total light received. Also, in urban areas the vast majority of the surfaces are asphalt and concrete, and such grey materials do not change significantly the spectral characteristics of the light.

As we have laid out, the correct calibration of nighttime images of the Earth from the ISS requires processing through a number of steps (Fig. 1). Whilst some of these undoubtedly have more significant effects on the final image than others, we recommend that all are carried out, as their influences on different images can vary greatly.

Calibration of nighttime images of the Earth taken from the ISS opens up enormous possibilities for studying the spatial occurrence of artificial lighting and how this is changing with time (nightly, through the year, and across years). Determining how the spatial and temporal variation covaries and determining the influences on other factors (e.g. human health and environmental impacts) is of great interest. As we have previously shown, by using colour-colour diagrams, inferences can also be drawn as to how differences and changes in the lighting technologies being used give rise to these spatio-temporal dynamics \citep{sanchezde2019colour}. 

\section{Acknowledgements}
We thank R. Moore of the Image Science and Analysis Group, NASA Johnson Space Center for information on ISS cameras and window properties. We thank Luc\'ia Garc\'ia for her help with improving some figures. We also thank the anonymous reviewers for their constructive feedback. This work was supported by the EMISSI@N project (NERC grant NE/P01156X/1), Fonds de Recherche du Qu\'ebec : Nature et Technologies (FRQNT), COST (European Cooperation in Science and Technology) Action ES1204 LoNNe (Loss of the Night Network), the ORISON project (H2020-INFRASUPP-2015-2), the Cities at Night project, FPU grant from the Ministerio de Ciencia y Tecnologia and F. Sánchez de Miguel.

Cameras were tested at Laboratorio de Investigaciónn Científica Avanzada (LICA), a facility of UCM-UPM funded by the Spanish program of International Campus of Excellence Moncloa (CEI). We acknowledge the support of the Spanish Network for Light Pollution Studies (MINECO AYA2011-15808-E) and also from STARS4ALL, a project funded by the European Union H2020-ICT-2015-688135. This work has been partially funded by the Spanish MICINN, (AyA2018-RTI-096188-B-I00), and by the Madrid Regional Government through the TEC2SPACE-CM Project (P2018/NMT-4291).

The ISS images are courtesy of the Earth Science and Remote Sensing Unit, NASA Johnson Space Center. Thanks to S. Doran for helping us to locate the missing image that we use as an example. We are very grateful to all members of the crews of the ISS from all agencies, NASA, ESA, JAXA, CSA-ASC and ROSCOSMOS, for their images.

\section*{Description of author's responsibilities}
A.S.M., J.Z., J.G., M.A., D.R.P., J.B. and K.J.G. conceived the study, A.S.M., M.A. and J. Z.  conducted the calibration procedures, A.S.M., F.O.  and J.Z. analyzed the results, D.R.P. acquired very many of the images used, and designed and developed the basic techniques for the acquisition of images and documented the ISS specifications, W.L.S. led the Earth Science and Remote Sensing Unit, NASA Johnson Space Center and provided key technical information, and A.S.M., K.J.G. and J.Z. wrote the original manuscript, A.S.M, J.Z., J.G., W.L.S., J.B. and K.J.G. conducted the funding requests. All authors reviewed the manuscript.

\section*{Declaration of interest}

A.S.M., J.Z. and K.J.G. are members of environmental organizations, including Bird Life, Celfosc and the International Dark-Sky Association.
A.S.M. occasionally provides consultancy advice for the Instituto de Astrofisica de Andalucia - CSIC and the company SaveStars Consulting S.L.

\bibliography{sample}

\begin{landscape}
\begin{table}[ht]
\begin{tabular}{lllllllll}
Model            & Kodak DCS760 & Nikon D1   & Nikon D2Xs & Nikon D3   & Nikon D3S   & Nikon D4    & Nikon D5              & Sony Alpha a7S II \\ \hline \hline
Agencies         & All          & All        & All        & All        & All         & All         & All                   & JAXA              \\ \hline
Model launch     & 22/03/2001   & 15/06/1999 & 01/06/2006 & 23/08/2007 & 4/10/2009   & 06/01/2012  & 05/01/2016            & 11/09/2015        \\ \hline
In use at ISS    & 2002–2007    & 2004–2006  & 2008–2014  & 2009–2012  & 2010-2015   & 2015–2017   & 2017-                 & 2018-             \\ \hline
Megapixels       & 6            & 3          & 13         & 13         & 13          & 17          & 21                    & 12                \\ \hline
Sensor format    & APS-H        & APS-C      & APS-C      & Full Frame & Full Frame  & Full Frame  & Full Frame            & Full Frame        \\ \hline
Sensor type      & CCD          & CCD        & CMOS       & CMOS       & CMOS        & CMOS        & CMOS                  & CMOS              \\ \hline
ISO Range        & 80 - 400     & 200 – 1600 & 100 – 800  & 200 -6400  & 200 – 12800 & 100 – 12800 & 100 – 102400          & 100 – 102400      \\ \hline
Image format     & RAW          & RAW        & RAW        & RAW        & RAW + TIFF  & RAW         & RAW                   & RAW               \\ \hline
Fitting          & Nikon F      & Nikon F    & Nikon F    & Nikon F    & Nikon F     & Nikon F     & Nikon F               & Sony E            \\ \hline
Amplification    & 1.3x         & 1.5x       & 1.5x       & 1x         & 1x          & 1x          & 1x                    & 1x                \\ \hline
Live view        & No           & No         & No         & Yes        & Yes         & Yes         & Yes                   & Yes               \\ \hline
Video resolution &              &            &            &            & 1280 x 720  & 1920 x 1080 & 3840 x 2160           & 4K                \\ \hline
Storage          & PCM CIA      & CF         & CF         & CF         & CF          & CF,XQD     & Dual CF,XQD & SD/SDH/C/SDXC\\ \hline    
\end{tabular}
\caption{Main characteristics of the most popular cameras used for nighttime photography and video at the ISS. Other cameras have been used but rarely or never for this purpose or only for specific experiments (like Nikon D800E and Nikon D850). More info at: https://eol.jsc.nasa.gov/FAQ. Source: https://www.dpreview.com/.}
\label{tab:1}
\end{table}

\end{landscape}

\begin{table}[ht]
\centering
\caption{Transmission coefficients for the most popular lenses used at the ISS or in light pollution research, relative to that of the 50mm f/1.4 lens. The values correspond to the ratio between the signal of surface brightness of lambertian reference source illuminated at 4 lux with a tungsten lamp, at 1/60 s and ISO 200 with a Nikon D5, for different lenses. This number is equivalent to the transmission stop (T-stop), but on a linear scale. Source: This work.}
\begin{tabular}{lcl}
\cline{1-2}
Lens                  & Transmission coefficient &  \\ \cline{1-2}
Nikon 180mm   f/2.8   & 0.297                    &  \\
Nikon 85mm   f/1.4    & 1.070                    &  \\
Nikon 85mm   f/1.8    & 0.765                    &  \\
Nikon 50mm   f/1.8    & 0.750                    &  \\
Nikon 50mm   f/1.4    & 1                        &  \\
Nikon 10.5mm   f/2.8  & 0.295                    &  \\
Nikon 24mm   f/1.4    & 1.050                    &  \\
Nikon 400mm   f/2.8   & 0.215                    &  \\
Sigma 8mm   f/3.5     & 0.187                    &  \\
Nikon 200mm   f/4     & 0.136                    &  \\
Nikon 28mm   f/1.4    & 1.630                    &  \\
Nikon 28mm   f/2.8    & 0.326                    &  \\
Nikon 28-70mm   f/2.8 & 0.324                    &  \\
Nikon 14-24mm   f/2.8 & 0.307                    &  \\ \cline{1-2}
\end{tabular}
\label{tab:2}
\end{table}

\begin{figure}[ht]
\centering
\includegraphics[width=\linewidth]{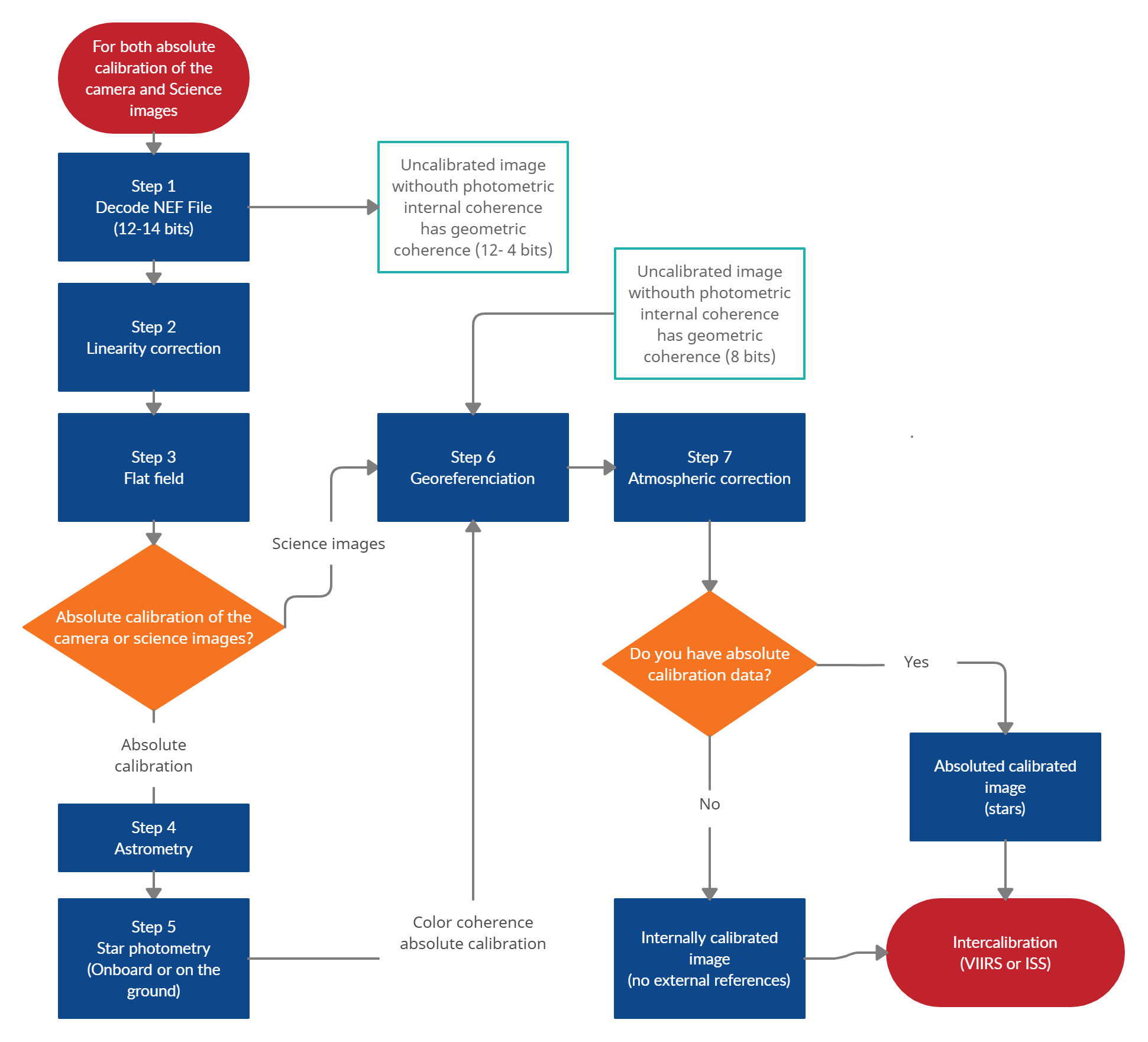}
\caption{Schema of the potential calibration paths. The red path requires the calibration of the camera using stars. This calibration provides colour coherence and absolute calibration based on stars. The black path indicates what can be done if we do not have an absolute calibrated image with stars, we can alternatively calibrate against the VIIRS-DNB. Some researchers have in the past used the blue path, using just decoded RAW images or even JPG images. We cannot recommend that procedure, as the images will not have internal photometric coherence or colour coherence, they will only have geometric coherence (the degree of incoherence will depend on the lens and dynamic range of the image).}
\label{fig:1}
\end{figure}

\begin{figure}[ht]
\centering
\includegraphics[width=\linewidth]{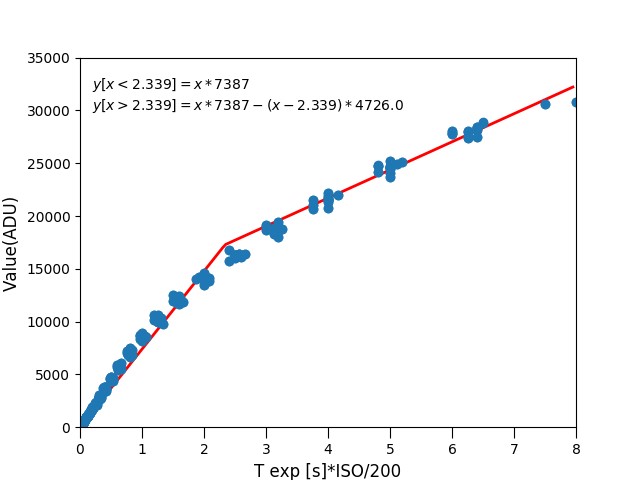}
\caption{Response of camera to a uniformly lit screen with stabilized source across all ISOs and times of exposure (T) in seconds. The dispersion of the dots is compatible with the errors of the stabilization (0.5\%). Response values expressed in dimensionless units (aka ADU). Data calculated with a D5, although compatible with previously calculated data for D3S. Source: this work.}
\label{fig:2}
\end{figure}

\begin{figure}[ht]
\centering
\includegraphics[width=0.5\linewidth]{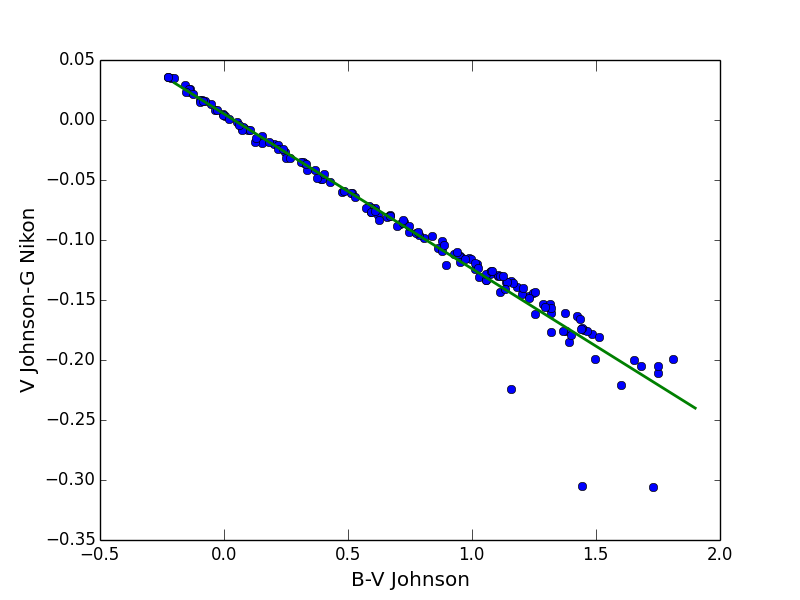}\includegraphics[width=0.5\linewidth]{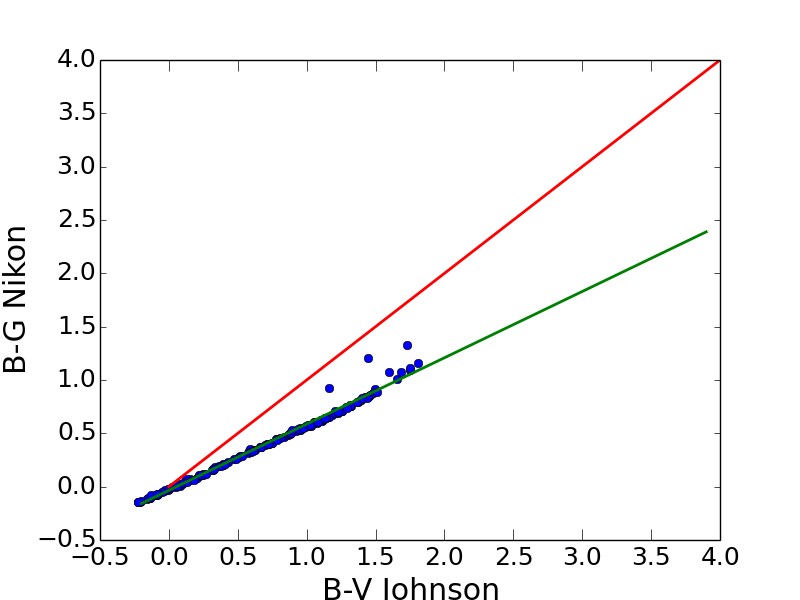}
\includegraphics[width=0.5\linewidth]{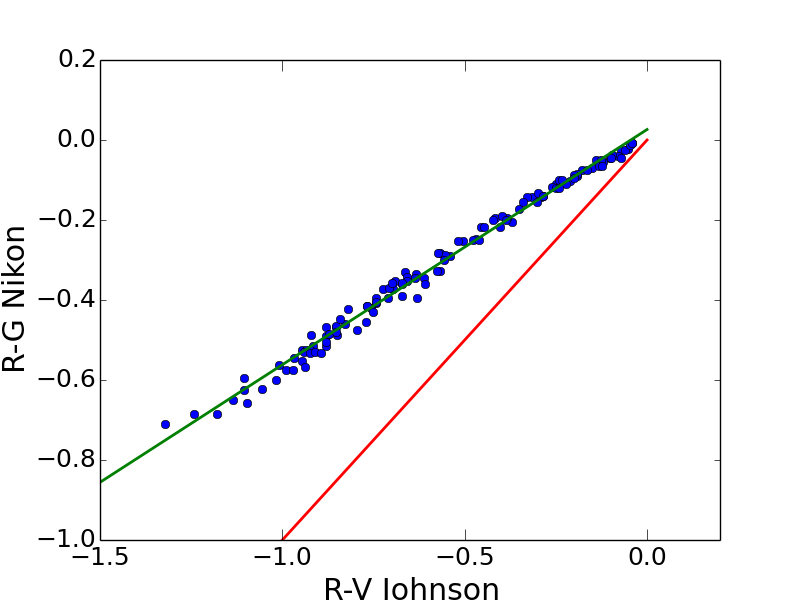}
\caption{Relationship between emissions in the RVB Johnson bands and the RGB Nikon bands. The 1:1 line is given in red, and the fit to the data in green.}
\label{fig:3}
\end{figure}

\begin{figure}[ht]
\centering
\includegraphics[width=0.5\linewidth]{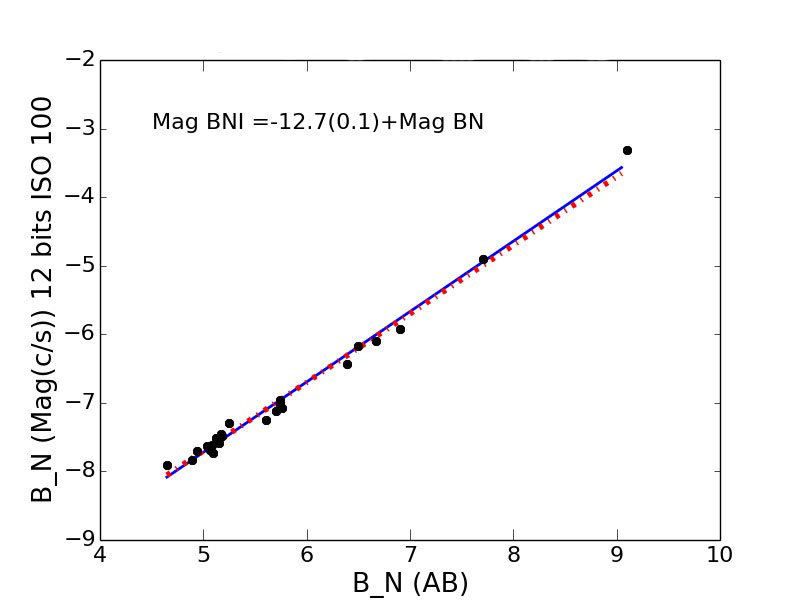}\includegraphics[width=0.5\linewidth]{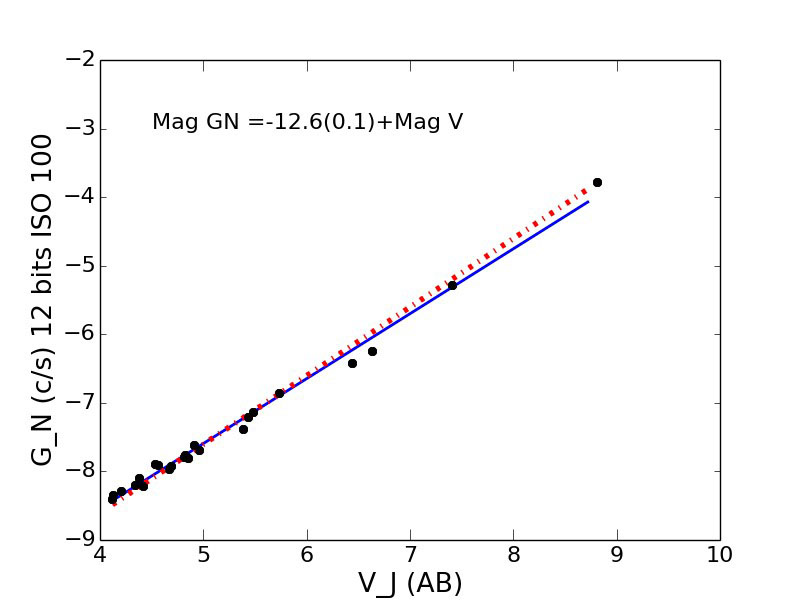}
\includegraphics[width=0.5\linewidth]{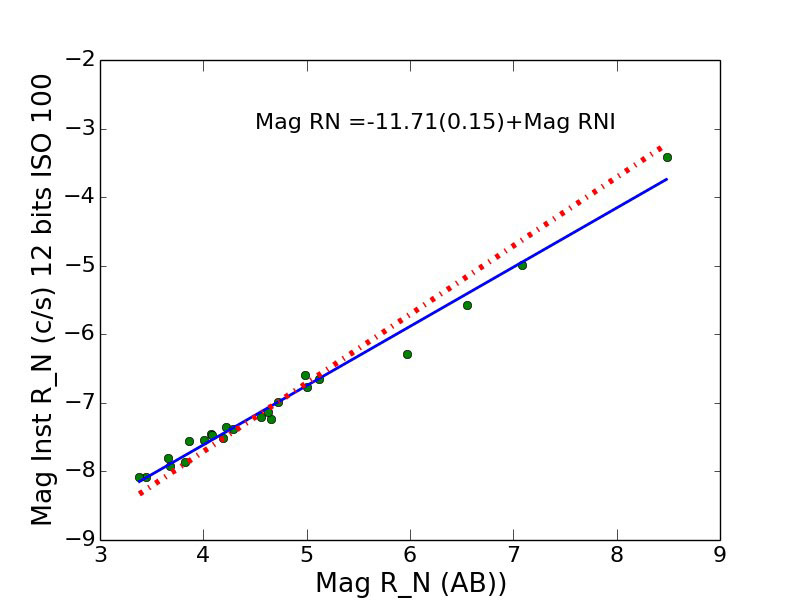}
\caption{Comparison of the radiance of stars in the Ducati II/237 catalogue with the values determined from calibration images taken with the D3S. Reference (X axis) brightness is in AB magnitudes. Instrumental magnitudes on Y axis. The measurements are in green, the blue line is the fit to all of the data, and the red line the fit excluding outliers using RANSAC. }
\label{fig:4}
\end{figure}

\begin{figure}[ht]
\centering
\includegraphics[width=1\linewidth]{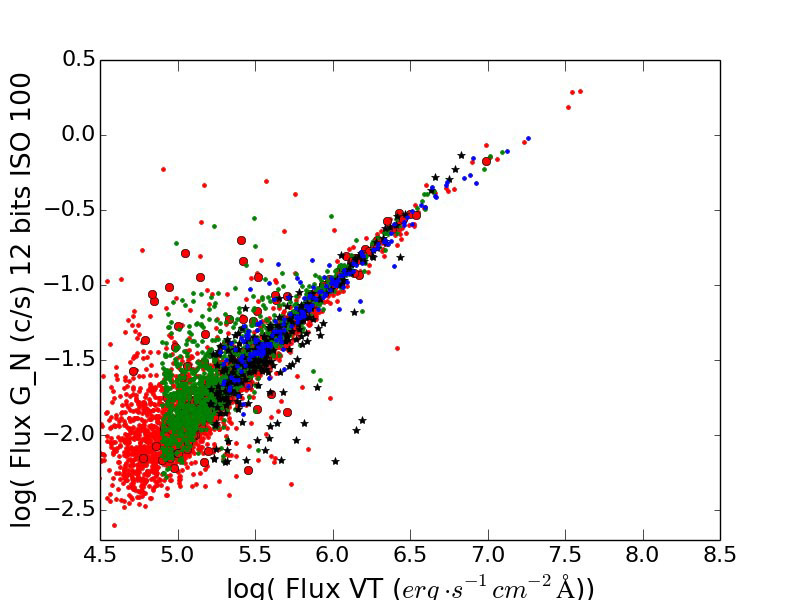}
\caption{Comparison of the radiance of stars in the Tycho I/259 catalogue with the values determined from calibration images taken with the D3S. Each colour represents a different image, which was obtained using different camera settings (exposure time, ISO). Longer exposure times and higher ISO lead to detection of more stars, but also present more saturation problems for bright stars, as exemplified by the image represented by the small red dots (more details in \cite{sanchezde2015variacion}).}  
\label{fig:5}
\end{figure}

\begin{figure}[ht]
\centering
\includegraphics[width=1\linewidth]{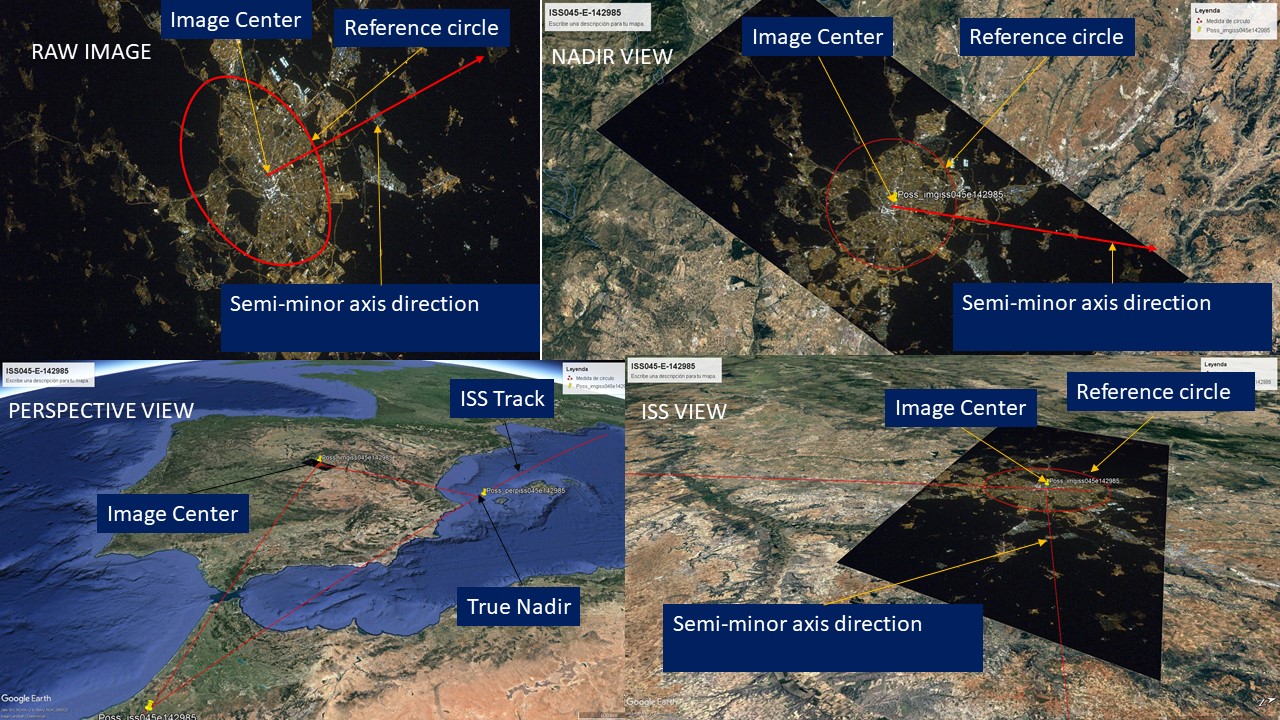}
\caption{With an ideal nadir view, that is typical of most observations from satellites, an imaginary circle seen from the nadir (top right view), would always be a circle on the ground. But, from a tilted observation, the circle would be transformed, approximately, into an ellipse (top left view and bottom right view), where the minor axis would be in the direction of observation and the major axis will point perpendicular to it. If we intersect view direction with the ground track of the ISS, we can find the true nadir point of observation (bottom left view). We have a first guess of the nadir point of observation thanks to the clock of the camera, but this can have an error as high as 500 km. The deformation is trivial to calculate, for example at the centre of the image, once the image has been georeferenced, using the reverse equations that give us the image coordinates from the ground equations. This simple technique allows us to find the true nadir. In a practical case, once the image is georeferenced, we do not use any target to stabilize the deformation of the image, we can use the georeferentiation formulas to estimate one at the centre of the image of the size that we prefer. In our case, we define a circle of 0.1 km of radius at the centre of the rectified image and we use the deformation formulas to calculate the deformed circle (ellipse).}
\label{fig:6}
\end{figure}

\begin{figure}[ht]
\centering
\includegraphics[width=1\linewidth]{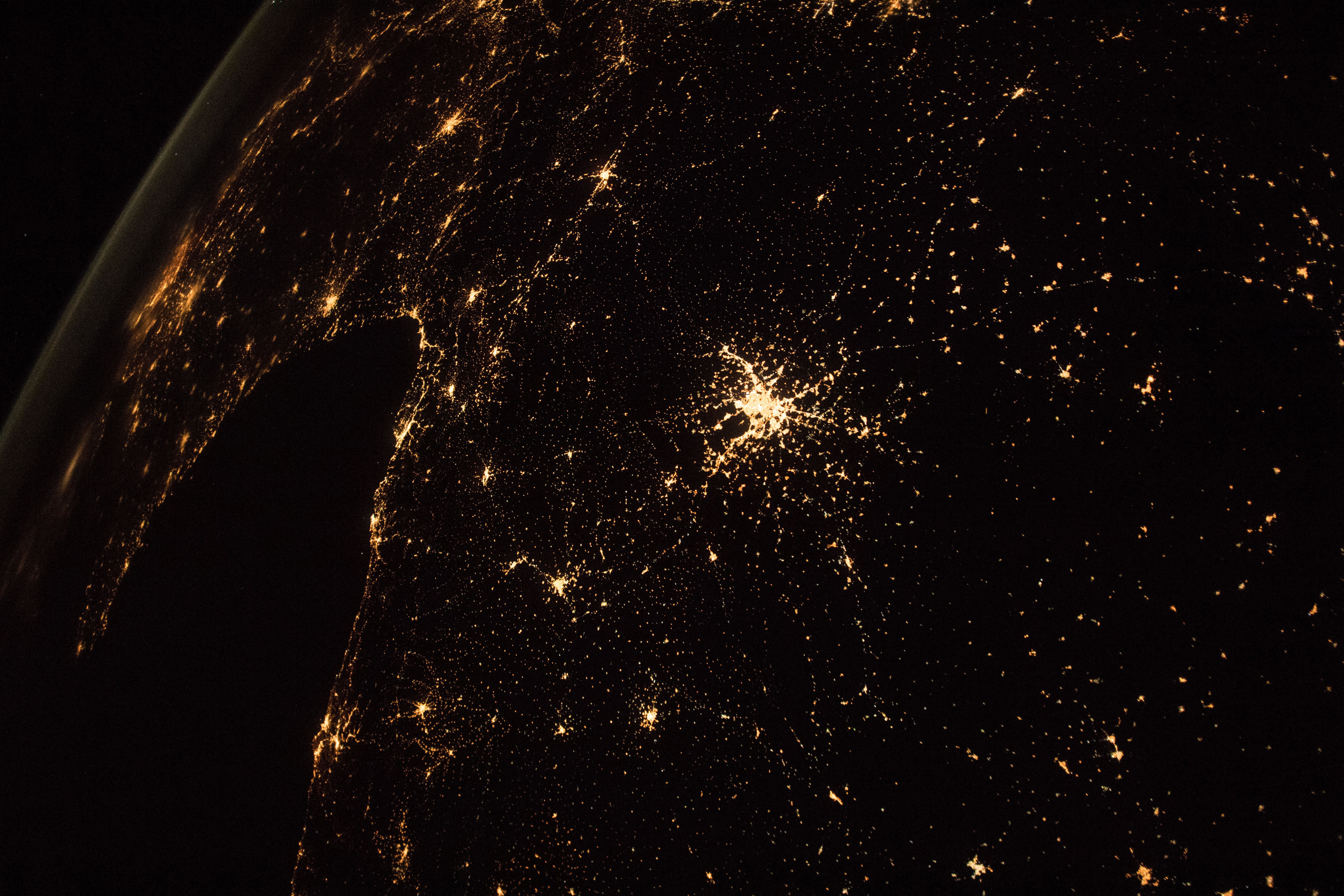}
\caption{Original JPG image of Spain used as an exemplar. Madrid (the spider-like lit area) looks to be saturated, although this is not actually the case. This is only one of the reasons not to use JPG images, although geometrically it is equivalent to the RAW image, colours, relative intensities, and other issues like gamma correction mean that this is not recommended.}
\label{fig:7}
\end{figure}

\begin{figure}[ht]
\centering
\includegraphics[width=0.6\linewidth]{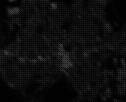}
\caption{Detail of the center of Madrid on  the RAW image of Spain, with the Bayer matrix and the 4 channels merged in one single image. The images do not appear clear because they have not been stretched and the different bands are entangled on the Bayer matrix, as mentioned in the Introduction. It is common for commercial software to use debayering algorithms that can produce artifacts and change the photometry, so we do not use that approach but calibrate each image separately.}
\label{fig:8}
\end{figure}

\begin{figure}[ht]
\centering
\includegraphics[width=1\linewidth]{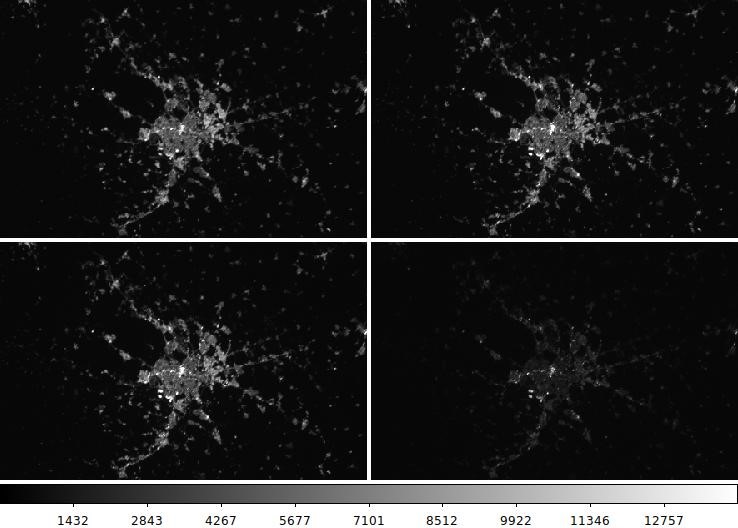}
\caption{Zoomed detail of Madrid from the exemplar image. The RAW image is extracted and separated into four different images. From top left to bottom right, Red (R1), Green (G2), Green (G3), and Blue (B4) channels.}
\label{fig:9}
\end{figure}

\begin{figure}[ht]
\centering
\includegraphics[width=1\linewidth]{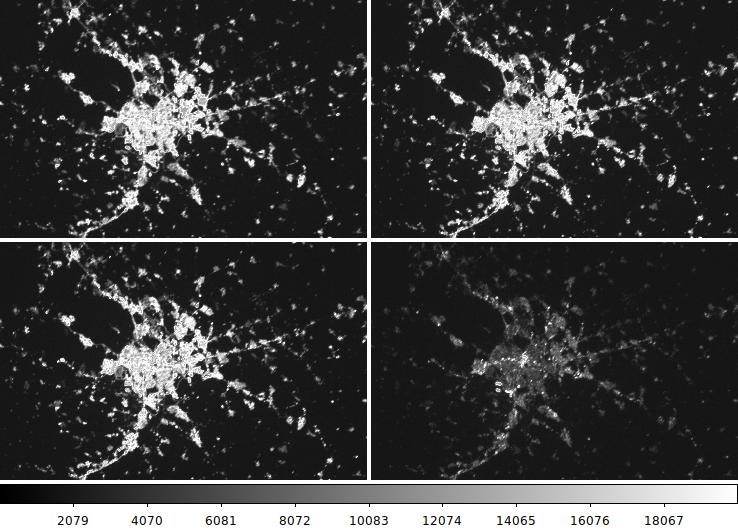}
\caption{Zoomed detail of the exemplar image, after linearity correction of the four separated channels.}
\label{fig:10}
\end{figure}

\begin{figure}[ht]
\centering
\includegraphics[width=1\linewidth]{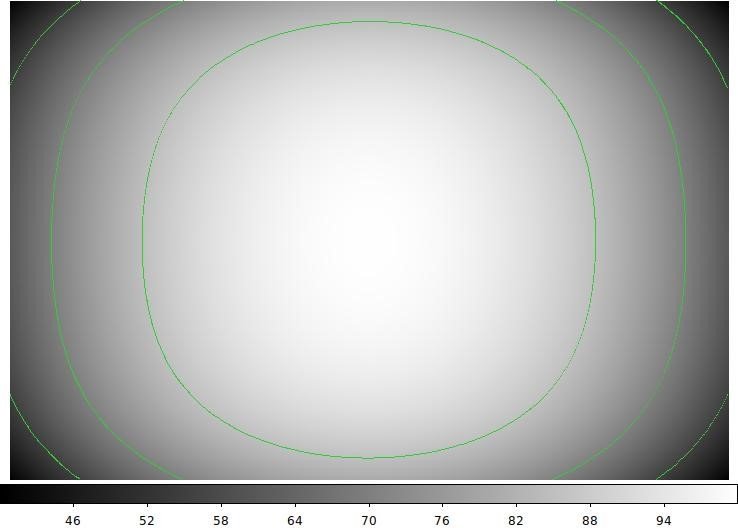}
\caption{Flat field correction for a 24 mm lens. The intensity at the centre corresponds to 100\%, the green lines 85\%, 45\% and 33\%, from inside to outside. Values presented correspond to percentage of the transmission compared with peak. }
\label{fig:11}
\end{figure}

\begin{figure}[ht]
\centering
\includegraphics[width=1\linewidth]{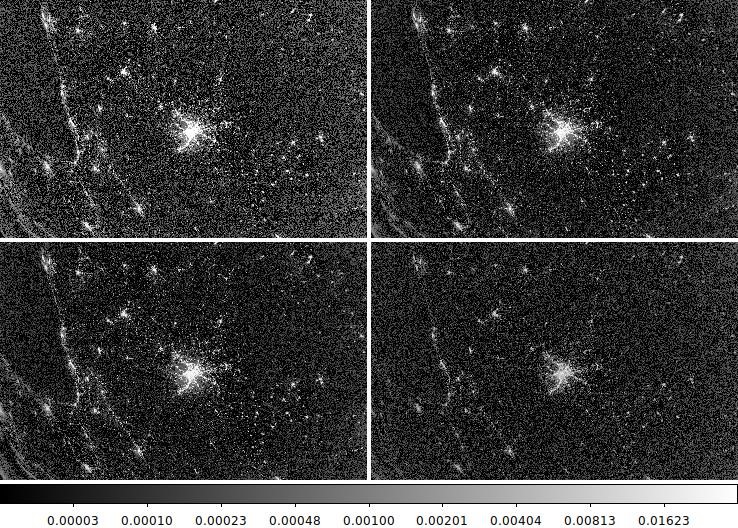}
\caption{Zoom in (not full picture) from fig 11 after the flat field correction. Results of photometric calibration of the exemplar image. The four extracted channels are now in values of $nW\cdot sr^{-1} cm^{-2} \si{\angstrom}^{-1}$. This image has not been stretched, so shows a lot of noise from low intensities and cosmic rays.}
\label{fig:12}
\end{figure}

\begin{figure}[ht]
\centering
\includegraphics[width=1\linewidth]{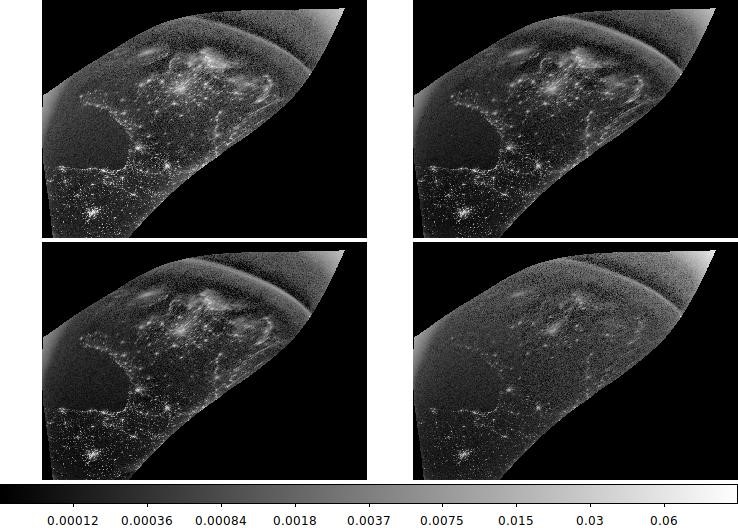}
\caption{Full picture rectified. The four channel images have to be georeferenced to create correspondence between pixel location and geographical location. This is the same image (Fig. 8), as before, but because the area of the pixels closer to the horizon is larger than for the pixels closer to the nadir, the centre of the image now is France, but this is just a perspective effect. The four extracted channels are now in values of $nW\cdot sr^{-1} cm^{-2} \si{\angstrom}^{-1}$. This image has not been stretched, so shows a lot of noise from low intensities and cosmic rays.}
\label{fig:13}
\end{figure}

\begin{figure}[ht]
\centering
\includegraphics[width=1\linewidth]{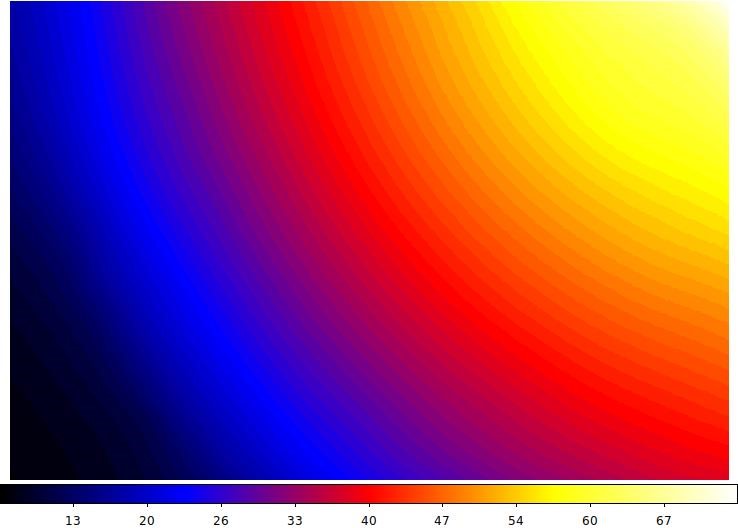}
\caption{Angular distance to the horizon for the exemplar image. The X and Y axis correspond to the coordinates of the image (see fig 13). These data, along with the altitude and the true nadir, are the basis of the atmospheric correction. }
\label{fig:14}
\end{figure}

\begin{figure}[ht]
\centering
\includegraphics[width=1\linewidth]{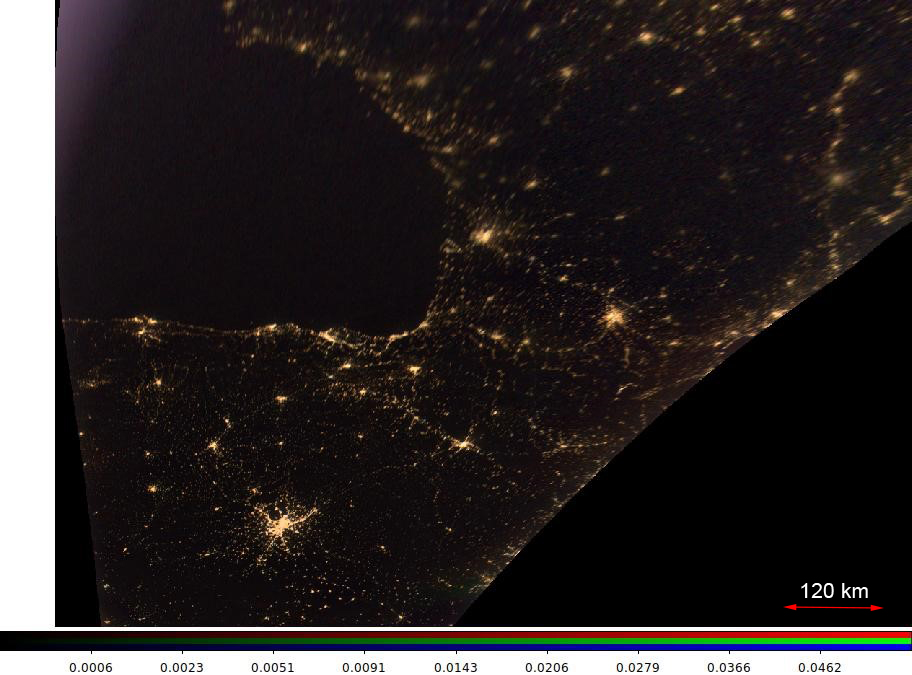}
\caption{Calibrated version of the exemplar image. Note that artefacts can appear at the edges because the flat field correction enhances noise in those areas. Also, the farther the pixels are from the nadir the blurrier they are. This is because pixels that are closer to the horizon correspond to larger areas than those closer to the nadir, so when the rectification takes place, there is less information in the first ones and errors propagate more than in the second ones. That is why, when several images are available of the same area, it is better to choose those that were acquired with longer focal lengths and closer to the nadir. Units in $nW\cdot sr^{-1} cm^{-2}$. This image has  been stretched, so does not show noise as clearly as in the other cases. }
\label{fig:15}
\end{figure}

\begin{figure}[ht]
\centering
\includegraphics[width=1\linewidth]{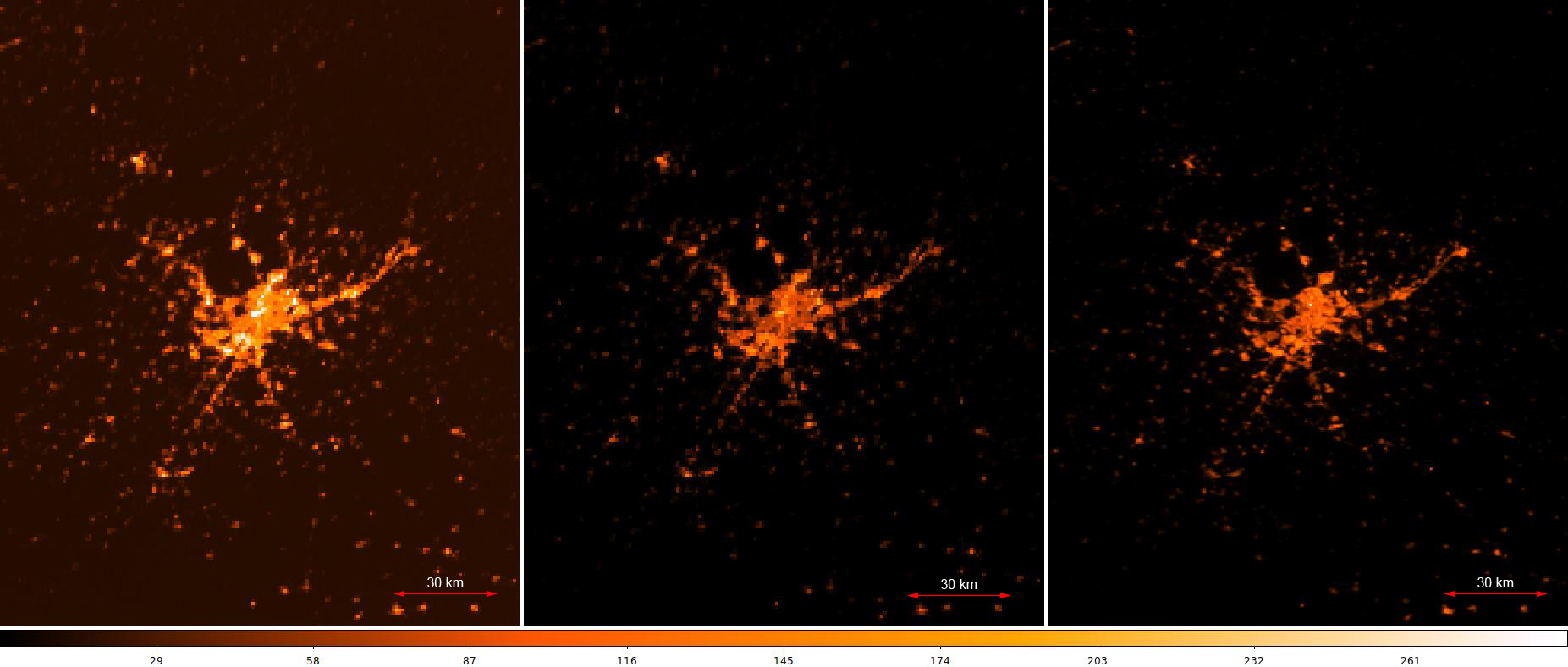}
\caption{Zoom to Madrid region. Left is the ISS Green band radiance calibrated with stars. Middle image is ISS green band inter-calibrated with VIIRS. Right image is VIIRS comparison image. Units in $nW\cdot sr^{-1} cm^{-2}$. The comparison is between the VIIRS October 2017 (November 2017 has a defect) image EOG average \cite{elvidge2013viirs} and ISS.}
\label{fig:16}
\end{figure}

\begin{figure}[ht]
\centering
\includegraphics[width=0.5\linewidth]{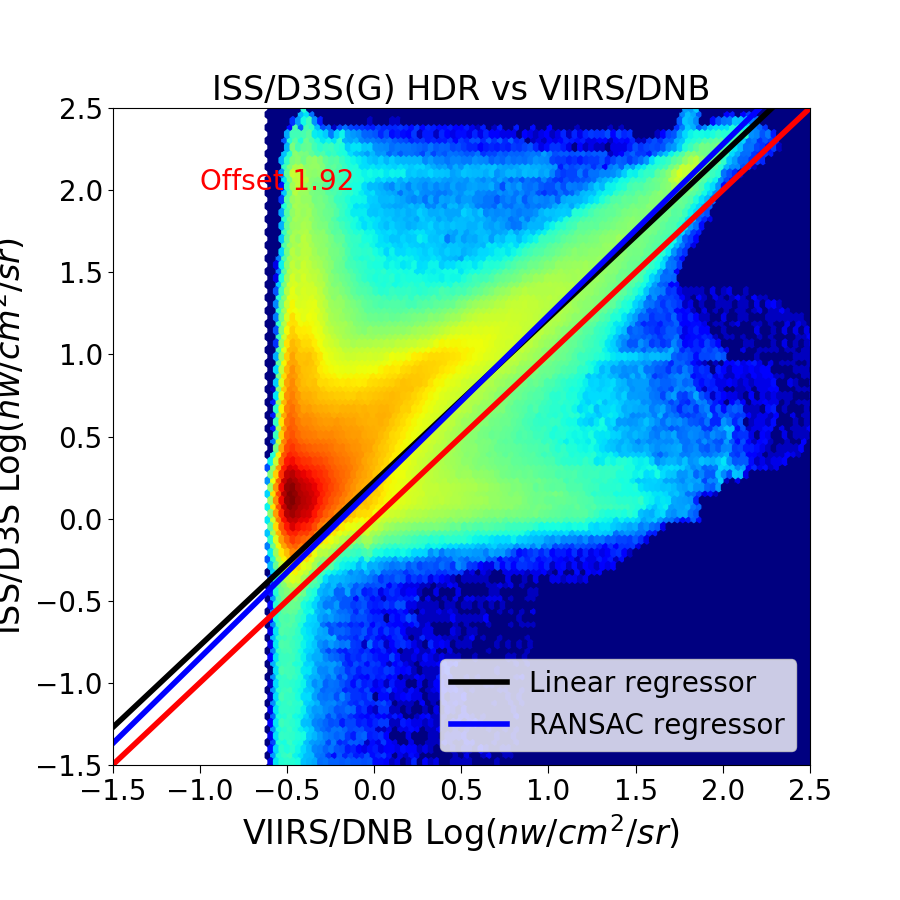}\includegraphics[width=0.5\linewidth]{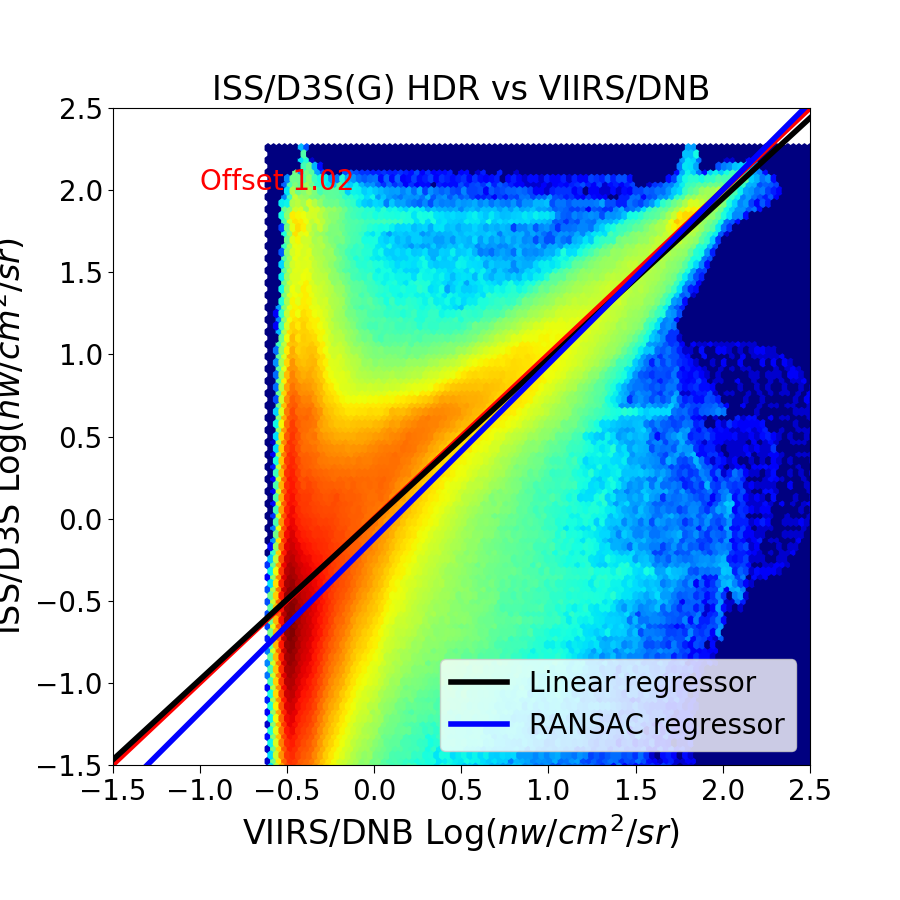}
\caption{Left: Comparison between VIIRS and ISS Green band star calibrated. Right: Comparison between ISS and VIIRS after background and slope offset. Red line corresponds to 1:1 relationship. Black line, standard linear fit, blue line RANSAC \cite{scikit-learn} linear fit. Slope Offset is calculated ratio from $100 nW\cdot sr^{-1} cm^{-2}$ ISS compared to corresponding VIIRS equivalent intensity. Background offset corresponds to maximum density points.  }
\label{fig:17}
\end{figure}

\begin{figure}[ht]
\centering
\includegraphics[width=1\linewidth]{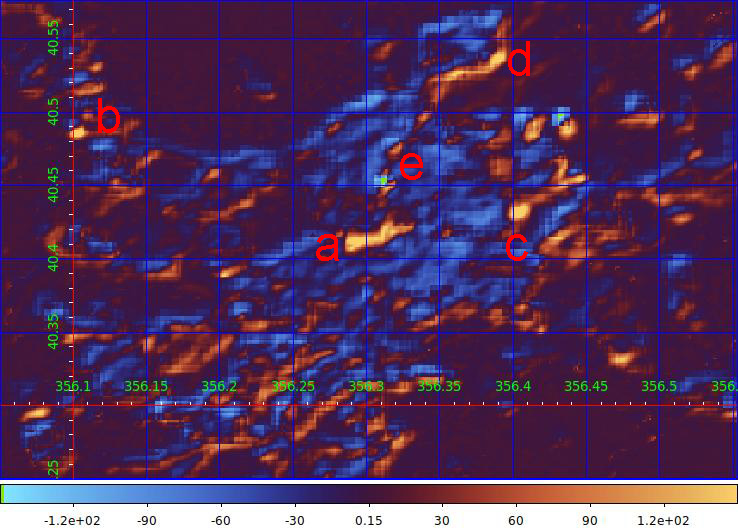}
\caption{Difference between the VIIRS and ISS image. The main differences are due to time of acquisition of the images. In the ISS image the lights can be seen of the main commercial centers of the Madrid area, the Sol square and surroundings (center of image (a)), the  commercial center Gran Plaza 2 (in the west(b)),  and the commercial center Plenilunio (in the East (c)) and Plaza Norte (in the north (d)). Also, there is a small difference in the georeferencing. The Santiago Bernabeu Stadium(e) does not appear on the ISS image, probably because of the tilt effect (dark area in the center of the image). Look-up table "Roma" from \cite{crameri_fabio_2021_4491293}.}
\label{fig:18}
\end{figure}

\begin{figure}[ht]
\centering
\includegraphics[width=1\linewidth]{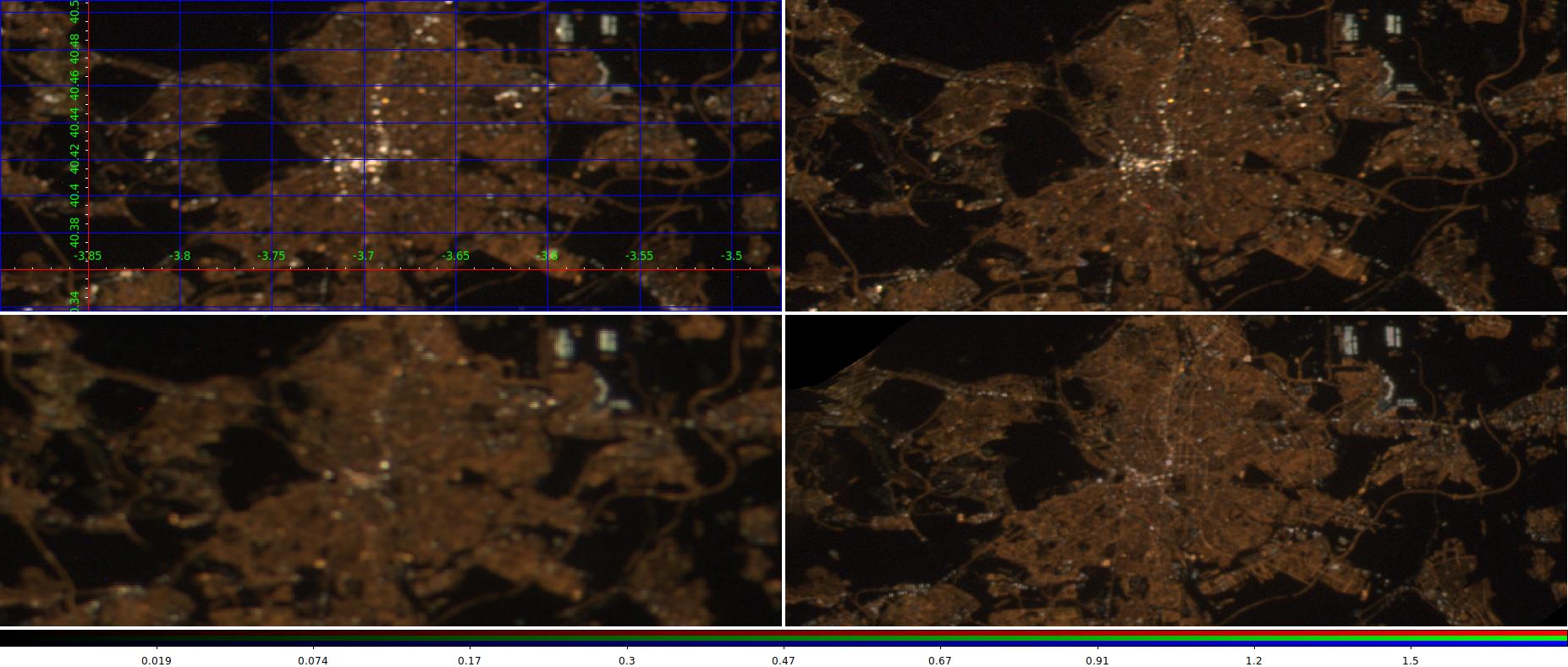}
\caption{ From left  to right and top to bottom, the images iss030e292893, iss030e292895, iss031e095634 and iss035e023371. Intensity scale in $nW\cdot sr^{-1} cm^{-2} \angstrom^{-1}$. The grid represents latitude and longitude.}
\label{fig:19}
\end{figure}

\begin{figure}[ht]
\centering
\includegraphics[width=1\linewidth]{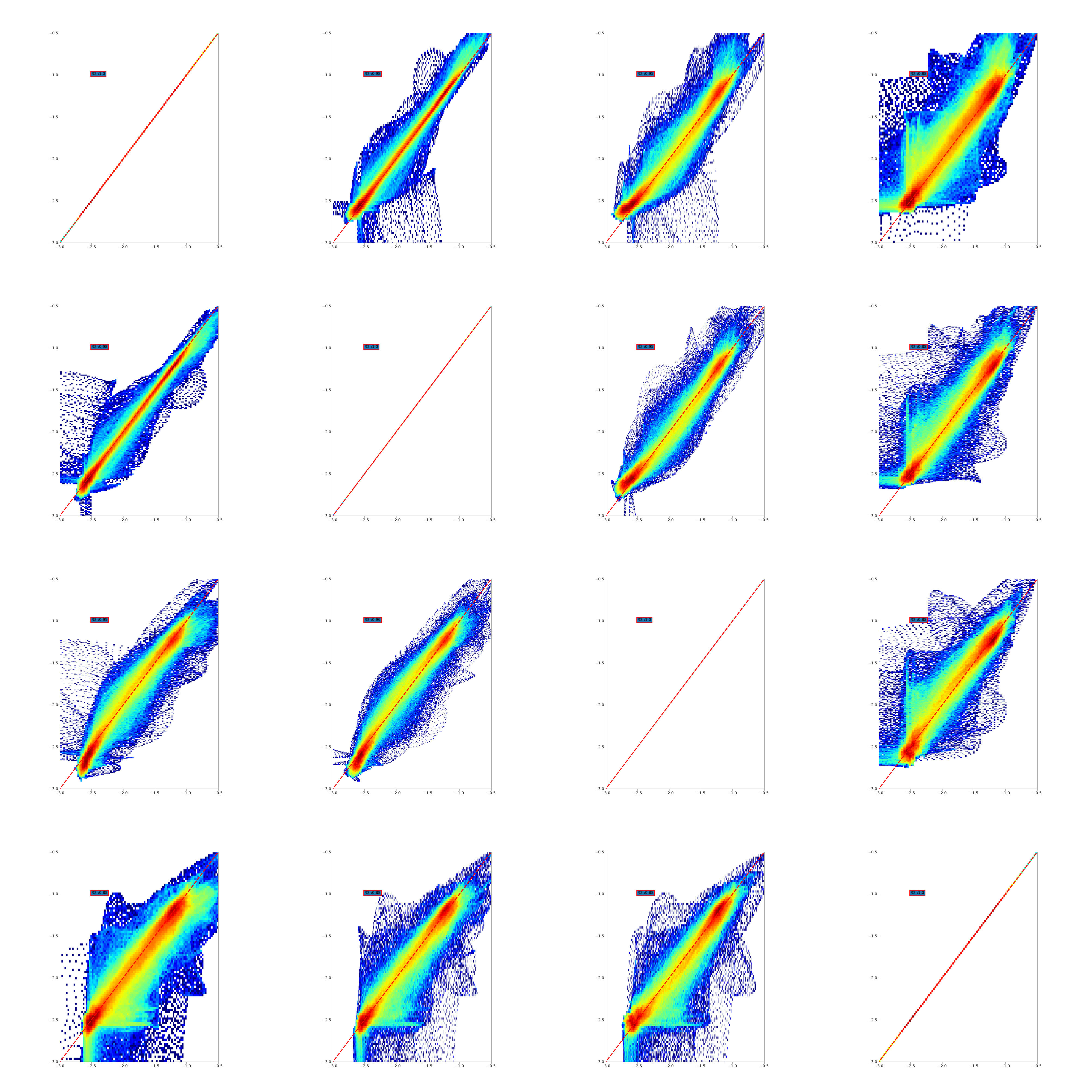}
\caption{ Density plots of the comparison between the images iss030e292893, iss030e292895, iss031e095634 and iss035e023371 on the Green band. The red lines indicate the 1:1 relationship. Units log10($nW\cdot sr^{-1}\cdot cm^{-2}\cdot \angstrom^{-1}$). }
\label{fig:20}
\end{figure}

\end{document}